\documentclass[prd,aps,floats,floatfix,eqsecnum,nofootinbib]{revtex4}
\usepackage{verbatim,graphicx,amssymb,amsbsy,bm,amsmath,rotating,epsfig}
\usepackage{psfrag}
\usepackage{hyperref}
\usepackage{fontenc}
\usepackage[latin1]{inputenc}
\usepackage{pstricks,pst-node,pst-text,pst-3d}
\usepackage{textcomp}
\newcommand{\be}{\begin{equation}}
\newcommand{\ee}{\end{equation}}
\newcommand{\bea}{\begin{eqnarray}}
\newcommand{\eea}{\end{eqnarray}}

\begin{document}
\title{Warm dark matter primordial spectra and \\ the onset of structure formation at redshift z }
\author{\bf C. Destri $^{(a)}$} \email{Claudio.Destri@mib.infn.it}
\author{\bf H. J. de Vega $^{(b,c)}$}
\email{devega@lpthe.jussieu.fr} 
\author{\bf N. G. Sanchez $^{(c)}$}
\email{Norma.Sanchez@obspm.fr} 
\affiliation{$^{(a)}$ Dipartimento di Fisica G. Occhialini, Universit\`a
Milano-Bicocca and INFN, sezione di Milano-Bicocca, Piazza della Scienza 3,
20126 Milano, Italia. \\
$^{(b)}$ LPTHE, Universit\'e Pierre et Marie Curie (Paris VI),
Laboratoire Associ\'e au CNRS UMR 7589, Tour 24, 5\`eme. \'etage, 
Boite 126, 4, Place Jussieu, 75252 Paris, Cedex 05, France. \\
$^{(c)}$ Observatoire de Paris,
LERMA. Laboratoire Associ\'e au CNRS UMR 8112.
 \\61, Avenue de l'Observatoire, 75014 Paris, France.}
\date{\today}
\begin{abstract}
Analytic formulas reproducing the warm dark matter (WDM) primordial spectra are obtained for
WDM particles decoupling in and out of thermal equilibrium; these formulas provide the
initial data for WDM non-linear structure formation. We compute and analyze the corresponding WDM overdensities
and compare them to the cold dark matter  (CDM) case.
We consider the ratio of the WDM to CDM primordial spectrum and the ratio of the 
WDM to CDM overdensities: they turn to be self-similar functions of $ k/k_{1/2} $
and $ R/R_{1/2} $ respectively, $ k_{1/2} $ and $ R_{1/2} $ being the wavenumber 
and length where the WDM spectrum and overdensity are one-half of the respective CDM magnitudes.
Both  $ k_{1/2} $ and $ R_{1/2} $ show scaling as powers of the WDM particle mass $ m $ while the 
self-similar functions are independent of $ m $.
The WDM primordial spectrum sharply decreases around $ k_{1/2} $
with respect to the CDM spectrum, while the WDM overdensity slowly decreases around  $ R_{1/2} $ 
for decreasing scales with respect to the CDM one.
The nonlinear regions where WDM structure formation takes place are shown and
compared to those in CDM: the WDM non-linear structures start to form later than in CDM,
and as a general trend, decreasing the DM particle mass delays the onset of the non-linear regime. 
The non-linear regime starts earlier for smaller objects than for larger ones;
smaller objects can form earlier both in WDM and CDM.
We compute and analyze the differential mass function $ dN/dM $ for WDM at redshift $ z $ 
in the Press-Schechter approach. The WDM suppression effect of small scale
structure {\bf increases} with the redshift $ z $.
Our results for $ dN/dM $ are useful to be contrasted with observations,
in particular for $ 4  \lesssim z \lesssim 12 $. We perfom all these studies for the most popular
WDM particle physics models. Contrasting them to observations should point out the precise {\bf value} 
of the WDM particle mass within the keV scale, and help to {\bf single out} the best WDM particle physics model.
\end{abstract}
\maketitle
\tableofcontents

\section{Introduction and Results}

Warm Dark Matter (WDM), that is dark matter formed by particles with masses in the keV scale
receives increasing attention today \cite{muchas}.

At intermediate scales $ \sim 100 $ kpc WDM provides the {\bf correct abundance} of substructures \cite{simuwdm}
and therefore WDM solves the CDM overabundance of structures for small scales.
For scales larger than $ ~ 100 $ kpc,
WDM yields the same results than CDM and agrees with all the observations:
small scale as well as large scale structure observations and CMB anisotropy observations.

Inside galaxy cores, below  $ \sim 100$ pc, $N$-body classical physics simulations 
do not provide the correct structures
for WDM because quantum effects are important in WDM at these scales.
Classical $N$-body WDM simulations exhibit cusps or small 
cores with sizes smaller than the observed cores \cite{coreswdm,mash}.
WDM predicts correct structures and cores with the right sizes for small scales (below kpc) 
when its {\bf quantum} nature is taken into account \cite{nos}.

{\vskip 0.2cm} 

A basic quantity in dark matter cosmology is the DM primordial power spectrum which
is obtained by solving the linearized Boltzmann-Vlasov equations till DM decouples. 

Although the  primordial power spectrum is obtained from the linearized Boltzmann-Vlasov equations,
it contains enough information to derive from it the mass function of formed structures in the Press-Schechter
approach as well as in the halo model and excursion set approaches.

We provide here analytic formulas that reproduce the primordial cosmological power for WDM.
Furthermore, we provide simple analytic formulas for the overdensity 
and the differential mass function of bounded structures as functions of the 
length scale and the redshift $ z $.

We perform all this study for four WDM fermion particle physics models where WDM
decouples in and out of equilibrium: Dodelson-Widrow, Shi-Fuller, 
$\nu$MSM and WDM thermal fermions. The first three models concern sterile neutrinos
while the last one can be applied to gravitinos.

{\vskip 0.2cm} 

In the thermal case, we obtain the WDM primordial power spectrum by introducing in the CAMB programmme
WDM fermions decoupling at thermal equilibrium. In the cases of decoupling out of thermal equilibrium,
we obtain the WDM primordial spectrum by solving the evolution Volterra integral equations 
we have derived in ref. \cite{fluc} from the linearized Boltzmann-Vlasov equations for WDM fermions.

{\vskip 0.1cm} 

\begin{figure}[h]
\begin{center}
\begin{turn}{-90}
\psfrag{"Rm.9.dat"}{Dodelson-Widrow}
\psfrag{"Rm.98.dat"}{Shi-Fuller}
\psfrag{"Rm1.3.dat"}{$\nu$MSM}
\psfrag{"Rm2.5.dat"}{FD Thermal}
\psfrag{"Rcdm.dat"}{CDM}
\includegraphics[height=13.cm,width=8.cm]{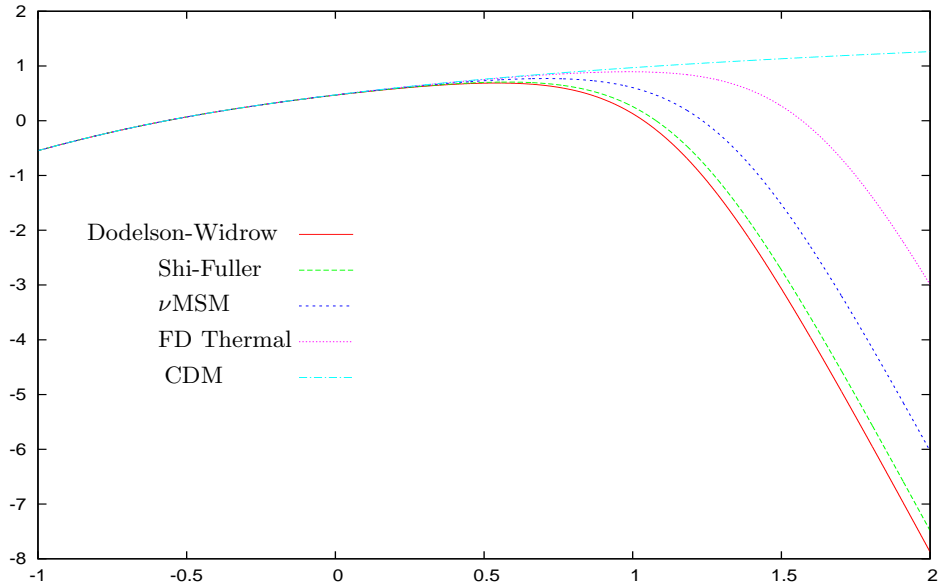}
\end{turn}
\caption{Warm dark matter primordial power spectrum $ \Delta_{wdm}^2(k) $
for a physical mass of 2.5 keV in four different WDM models
and the CDM primordial power spectrum $ \Delta_{cdm}^2(k) $.
We plot the ordinary logarithm of the power spectrum $ \Delta^2(k) $
vs. the  ordinary logarithm of $ k \; $ Mpc$/h$.}
\label{espprim}
\end{center}
\end{figure}

We depict in fig. \ref{espprim} the WDM primordial power spectrum for a physical 
WDM mass of 2.5 keV in four different WDM models, and the CDM primordial power spectrum.
We choose the value 2.5 keV because astrophysical evidences from galaxies points towards
a dark matter particle mass $ m $ around this value \cite{muchas}. In any case, it is easy to compute
the  WDM primordial power spectrum for other values of $ m $ in the keV scale by using 
the analytic formulas eqs.(\ref{trafu})-(\ref{valores}) provided below.

{\vskip 0.1cm} 

The ratio of the WDM and CDM primordial power spectra
$$
 T^2(k) \equiv \frac{\Delta_{wdm}^2(k)}{\Delta_{cdm}^2(k)} \; ,
$$
turns to exhibit the scaling form
\be\label{funsca}
T^2(k) = \Phi\!\left(\frac{k}{k_{1/2}} \right) \quad , 
\ee
where the WDM primordial power is one half of the CDM primordial power
at the wavenumber $ k_{1/2} $ 
$$ 
T^2(k_{1/2}) = 1/2 \; .
$$

The CAMB results we obtain are appropiately reproduced by the analytic formula
\be\label{funs}
\Phi(x) = \frac1{\left[1+\left(2^{1/b}-1\right) \; x^a \right]^b} \quad , \quad 2^{1/b}-1=0.167\quad ,
\ee
where $ a = 2.304 \quad , \quad b = 4.478 \; , $ 
are independent of the WDM particle mass $ m $. We see from fig. \ref{fitpot} 
that eqs.(\ref{funsca})-(\ref{funs}) provide an excellent fit for $ T^2(k) $ computed with CAMB.

{\vskip 0.1cm} 

The characteristic scale $ k_{1/2} $ at which $ T^2(k_{1/2}) = 1/2 $ does depend on $ m $ as
$$ 
k_{1/2} = 6.72 \; \left(\frac{m_{FD}}{\rm keV}\right)^{1.12} \; \frac{h}{\rm Mpc} \; .
$$
Here $ m_{FD} $ is the WDM particle mass for fermions decoupling at thermal equilibrium.

WDM particles decoupling ultrarelativistically and out of thermal equilibrium
in different WDM particle models behave just as if their masses were different \cite{fluc}.
The masses of WDM particles in different models which give the same
primordial power spectrum are related according to the formula \cite{fluc}
\be\label{equi}
m_{DW} \simeq 2.85 \; {\rm keV} \; \left(\frac{m_{FD}}{\rm keV}\right)^{4/3} \quad , \quad 
m_{SF} \simeq 2.55 \;  m_{FD}  \quad , \quad 
m_{\nu{\rm MSM}} \simeq 1.9  \; m_{FD} \; ,
\ee
FD stands for WDM fermions decoupling in thermal equilibrium (Fermi-Dirac).
DW, SF and $\nu$MSM  stand for WDM sterile neutrinos decoupling out of thermal equilibrium 
in the Dodelson-Widrow\cite{dw}, Shi-Fuller\cite{sf} and $\nu$MSM \cite{st} models, respectively.

These relations ensure that the density and anisotropic stress
fluctuations of WDM and neutrinos are identical in the coupled evolution Volterra equations derived
in ref. \cite{fluc}. Therefore, the primordial WDM spectrum is the same for thermal fermions
and out of equilibrium sterile neutrinos when eqs.(\ref{equi}) hold.

Table \ref{valK} displays WDM particle masses providing the same primordial power 
in four different WDM particle models.

{\vskip 0.2cm} 

We study the expected overdensity $ \sigma(M,z) $ of mass $ M $ over a 
comoving radius $ R $ at redshift $ z $ in the linear regime. The overdensity is given by
$$
\sigma^2(R,z) = \int_0^{\infty} \frac{dk}{k} \; \Delta^2(k,z) \; W^2(kR) \quad 
{\rm where} \quad M(R) = \frac43 \; \pi \; \Omega_{dm} \; \rho_c \;  R^3 \; ,
$$
$ \rho_c $ is the critical density of the Universe and $ W(x) $ is the 
window  function:
$$
W(x)  =\frac3{x^3} \; (\sin x - x \; \cos x) \; .
$$
We plot in fig. \ref{sigma} the expected overdensity $ \log_{10} \sigma^2(R,z=0) $ vs.  
$ \log_{10} [R \; h / {\rm Mpc}] $ for a $ m =2.5 $ keV WDM particle in four different WDM 
fermion models and in CDM.

The WDM and CDM primordial power spectra as well as the overdensities depend 
on the redshift as \cite{wein,gor}
\be\label{Delz}
\Delta^2(k,z) = \frac1{(z+1)^2} \; \frac{g^2(z)}{g^2(0)} \; \Delta^2(k,0) \; ,
\ee
where $ g(z) $ is the function that provides the suppression of the growth of matter fluctuations 
due to the cosmological constant. The function $ g(z) $ is explicitly given by eq.(\ref{gz}).

{\vskip 0.1cm} 

We introduce the relative overdensity $ D(R) $
$$
 D(R) \equiv \frac{\sigma^2_{WDM}(R,z)}{\sigma^2_{CDM}(R,z)} \; ,
$$
which turns out to be $ z $ independent due to eq.(\ref{Delz}).
In fig. \ref{sigD}  we plot $ \log_{10} D(R) $ vs. $ \log_{10} [R \; h/{\rm Mpc}] $ 
for a particle mass $ m =2.5 $ keV in four different WDM models.

{\vskip 0.1cm} 

We define the length scale $ R_{1/2} $ where the WDM overdensity is one-half of the CDM overdensity:
$$
 D(R_{1/2}) = \frac12 \; ,
$$
and find for fermionic WDM particles of mass $ m_{FD} $ decoupling at thermal equilibrium:
\be
R_{1/2} = 73.1 \; \frac{\rm kpc}{h} \; \left(\frac{\rm keV}{m_{FD}}\right)^{1.45}
\ee
For WDM particles decoupling out of thermal equilibrium, eq.(\ref{noneq}) allows to make the appropriate
conversions. 

{\vskip 0.1cm} 

Structure formation is therefore suppressed in WDM with respect to CDM for scales below $ R_{1/2} $.
For scales larger than $ R_{1/2} $, structure formation in WDM and CDM are similar.

{\vskip 0.1cm} 

The relative overdensity $ D(R) $ vanishes in the limit $ R \to 0 $ 
expressing the suppression of small scale structures in WDM,
and tends to unity for large scales where WDM and CDM exhibit identical behaviour. 
More precisely, we find
\be\label{limiI}
D(R)\buildrel{R \ll R_{1/2}}\over= c_1 \; \left(\frac{R}{R_{1/2}}\right)^{0.37} \quad , \quad  
D(R)\buildrel{R \gg R_{1/2}}\over=1 - c_2 \; \left(\frac{R_{1/2}}{R}\right)^2 \; .
\ee
where $ c_1 $ and $ c_2 $ are numerical constants.

{\vskip 0.1cm} 

The relative overdensity $ D(R) $ turns to exhibit a scaling form
$$
 D(R) = \Psi\!\left(\frac{R}{R_{1/2}} \right) \quad , 
$$
where the function $ \Psi(x) $ is independent of the WDM particle mass $ m $.
$ \Psi(x) $ can be reproduced by the simple formula
\be\label{Dfit}
 \Psi(x) = \frac1{\left[1 + \left(2^{1/\beta}-1\right) \; x^{-\alpha}\right]^\beta}
\ee
which is the configuration space analogue of eq.(\ref{funs}) for the power spectrum and
which reproduces the limiting behaviours in eqs.(\ref{limiI}).
The best fit is obtained for the values
$$
\alpha \simeq 2.2 \quad , \quad  \beta \simeq 0.17 \quad ,\quad 2^{1/\beta}-1 \simeq 58 \; .
$$

While the transfer function $ T^2(k) $ {\bf sharply} decreases in wavenumber space around $ k_{1/2} $
as $ k^{a \, b} $ with a {\bf large} exponent $ a \, b  = 10.3 $ in eq.(\ref{funs}), 
the relative overdensity $ D(R) $ {\bf slowly} decreases 
for decreasing scales $ R \sim R_{1/2} $ as $ R^{\alpha \, \beta} $
with a {\bf small} exponent $ \alpha \, \beta= 0.37 $ in eq.(\ref{Dfit}).

\bigskip

In fig. \ref{nolin} we display the lines where $ \sigma^2(M,z) = 1 $,
in the $ z, \; \log [ h \; M/{\rm Mpc}] $ plane 
for $ m =2.5 $ keV in four different WDM models and in CDM. 
Above these lines we have the linear regime where $ \sigma^2(M,z) < 1 $
and below these lines we have the nonlinear regime where $ \sigma^2(M,z) > 1 $.

We see from fig. \ref{nolin} that in WDM the linear regime lasts longer than in CDM.
That is, for a given object of mass $ M $, the nonlinear regime in WDM starts later than in CDM, 
i. e. non-linear structures start to form later in WDM than in CDM. 

In addition, in WDM, for a given particle physics model, the smaller is the particle mass,
the later the nonlinear regime starts.

We see as a general trend that decreasing the DM particle mass delays the onset of the nonlinear
regime, the earlier nonlinear onset occurring in the extreme case of CDM particles 
with masses $ \sim $ GeV.

Fig. \ref{nolin} shows the redshift where the nonlinear regime starts for the different
WDM particle physics models and a WDM particle mass $ m =2.5 $ keV.

The non-linear regime starts earlier for smaller objects than for larger ones.
Smaller objects can form earlier. This suggests that regimes of hierarchical structure formation
are present in WDM as confirmed by N-body simulations \cite{mash}.

\bigskip

We see from figs. \ref{SMz} that the number of structures formed in WDM at small scales
is smaller than in CDM. As expected, the suppression of structure formation is
larger the smaller is the mass and size of the structures.

For large scales, $ M \gtrsim 10^{11} \; M_\odot $ , both CDM and WDM give the same results.

\medskip

Moreover, the suppression effect on the number of structures 
{\bf increases} with the redshift $ z $. Let us define the suppression function 
$ \beta_z(M) $ in terms of the ratio between
the WDM and CDM differential mass functions for mass $ M $ and redshift $ z $
\be\label{defbez}
\beta_z(M) \equiv  1 - \frac{S_{WDM}(M,z)}{S_{CDM}(M,z)} \; .
\ee
In particular, the extreme cases being:
\begin{itemize}
\item{$ \beta_z(M) = 1 $ total suppression of structures in WDM: $ S_{WDM}(M,z) = 0 $.}
\item{$ \beta_z(M) = 0 $ no suppression of structures in WDM: WDM and CDM giving
identical results.}
\end{itemize}
We display in Table \ref{zalto} the values of $ \beta_z(M) $ for $ z = 0 $ and
$ z = 10 $. We choose for WDM the Dodelson-Widrow model. For the other WDM particle models 
we find similar results.

{\vskip 0.1cm} 

Table \ref{zalto} shows that for fixed redshift the suppression function $ \beta_z(M) $ decreases for
increasing $ M $ and that for fixed $ M $ it increases with the redshift. Namely, 
the suppression of structure formation in WDM compared with CDM, {\bf increases} with the redshift $ z $.

{\vskip 0.1cm} 

The suppression function $ \beta_0(M) $ at $ z = 0 $, varies from 99.1\% for masses $ 10^6 \; M_\odot/h $
to 34.8\% for masses $ 10^{10}\; M_\odot/h $.
The suppression function is {\bf higher} at $ z = 10 $ where it varies from 99.5\% to 51.8 \%
in such range of galaxy masses (Table \ref{zalto}). We see that 
for galaxy masses $ M $ about $ 10^6 \; M_\odot/h $
the suppression of structure formation in WDM does not practically change with the redshift in
the interval $ 0 < z < 10 $.
For large masses $ M = 10^{10}\; M_\odot/h $, the suppression effect at
$ z = 0 $ is 67.3\% of the suppression at $ z = 10 $ (Table \ref{zalto}).

\bigskip

This suppression of small scale structures in WDM at large redshifts reduces
star formation. Decay of WDM sterile neutrinos in X-rays can boost the 
production of molecular hydrogen and speed up early star formation \cite{star}.
Star formation in WDM can take place at redshifts $ \sim 100 $ \cite{star}.
WDM plus baryons simulations produce baryon filaments of free 
streaming length size which by fragmentation lead to star formation 
through multiple stellar cores \cite{gaoth}. 
This additional mode of star formation through filaments 
(absent in CDM) could compensate the decrease of the number
of DM halos that host early-star formation \cite{gaoth}. Recent Herschel 
observations point towards star formation through filaments \cite{herschel}.
Recent WDM simulations claim that star formation reduces in WDM
only for WDM particles lighter than 2 keV \cite{herpi}.

\bigskip

Our theoretical predictions are therefore useful to compare with 
the available observations at large $ z $. Interestingly enough, besides the fact that WDM reproduces 
the observed structures better than CDM \cite{mfl},  contrast of these predictions
to observations should point out the
{\bf value} of the WDM particle mass, and help to {\bf single out} the best WDM particle physics
model.

\section{Primordial Cosmological Power Spectrum}\label{sec2}

The primordial matter cosmological power spectrum in dimensionless form is given by \cite{dod}
\be
\Delta^2(k) = \frac{k^3}{2 \, \pi^2} \; P(k)
\ee
This quantity gives the anisotropy of the CMB fluctuations for large scales
beyond the Mpc as well as the seeds for the structure formation (galaxies and cluster of galaxies) 
 in the universe.
It is therefore important to dispose of practical formulas for $ \Delta^2(k) $
at all scales which can be used as the starting point of the nonlinear gravitational
evolution of structures.

{\vskip 0.2cm} 

The dependence of $ \Delta^2(k,z) $ on the redshift $ z $ factors out as \cite{wein,gor}
\be\label{delz}
\Delta^2(k,z) = \frac1{(z+1)^2} \; \frac{g^2(z)}{g^2(0)} \; \Delta^2(k) \; ,
\ee
where $ \Delta^2(k) \equiv \Delta^2(k,0) $, and
$ g(z) $ is the function that provides the suppression of the growth of matter fluctuations 
due to the cosmological constant. The function $ g(z) $ can be written as
\be\label{gz}
g(z) = \frac56 \; \int_0^1 \frac{dt}{(1-t)^\frac16 \; (1+t \; \omega \; a^3)^\frac13}=
{}_2F_1\left(\frac13,1;\frac{11}6; -\omega \; a^3\right) \; ,
\ee
where $ {}_2F_1\left(\alpha,\beta;\gamma;z\right) $ is the hypergeometric function
and $ a $ stands for the scale factor,
$$
\omega\equiv \frac{\Omega_\Lambda}{\Omega_m} \simeq 2.7 \quad ,
\quad a = \frac1{z+1} \; .
$$
Notice that 
$$ 
g(0) = 0.760188\ldots \; .
$$

{\vskip 0.2cm} 

We approximate $ g(z) $ by the polynomial expression
\be\label{fgz}
g_P(z)= 0.99997 + A \; a^3 + B \; a^6 + C \; a^9 \quad {\rm where} \quad
A =-0.47666697 \; , \;  B =0.41261397  \; , \;  C =-0.17873702 \; .
\ee
This polynomial approximation to $ g(z) $ differs from the exact analytic expression eq.(\ref{gz})
by less than 0.003 for all positive $ z $. 

{\vskip 0.2cm} 

It is useful to quantitatively identify the effects due to the warm dark matter
in the primordial power spectrum $ \Delta^2(k,0) $.
At large scales, the primordial power spectrum is the same
for CDM and WDM. At small scales the WDM power spectrum is suppressed with respect to the
CDM  power spectrum. The characteristic scale of suppression is the free streaming length
(similar to the Jeans' length) which for WDM particles decoupling ultrarelativistically 
is given by \cite{fluc}. 
\be\label{lfs}
l_{fs} = 210 \, {\rm kpc} \; \frac{\rm keV}{m_{FD}} \; \left(\frac{100}{g_d}\right)^\frac13   \; ,
\ee
$ g_d $ being the effective number of UR degrees of freedom at the DM decoupling.

\medskip

It is convenient to consider the transfer function ratio $ T(k) $ defined by
\be\label{trafu}
 T^2(k) \equiv \frac{\Delta_{wdm}^2(k)}{\Delta_{cdm}^2(k)} \; .
\ee
For large scales $ k \ll 1/l_{fs} $, this transfer function tends to unity while
it vanishes for small scales $ k \gg 1/l_{fs} $.

\bigskip

For WDM decoupling at thermal equilibrium,
we have computed $ T^2(k) $ by introducing in the CAMB programmme
WDM fermions decoupling at thermal equilibrium for various values of the WDM particle
mass $ m $. (For informations about CAMB see ref. \cite{camb}).
For WDM decoupling out of
thermal equilibrium we used the evolution Volterra integral equations derived in ref. \cite{fluc}. 

\medskip

The results we obtained from CAMB for $ T^2(k) $ can be conveniently fitted with the simple formula:
\be\label{fitt2}
T^2(k) = \frac1{\left[1+  \displaystyle \left(\frac{k}{\kappa}\right)^a\right]^b}
\ee
We find that the exponents $ a $ and $ b $ are independent of the WDM particle
mass $ m $ while the coefficient $ \kappa $ scales with  $ m $.  In our best fit:
\be\label{valores}
a = 2.304 \quad , \quad b = 4.478 \quad , \quad 
\kappa = 14.6 \; \left(\frac{m_{FD}}{\rm keV}\right)^{1.12} \; \frac{h}{\rm Mpc} \; ,
\ee
where $ m_{FD} $ is the WDM particle mass for fermions decoupling at thermal equilibrium.
WDM particles decoupling out of equilibrium are discussed in sec. \ref{secnoe}.

\begin{figure}[h]
\begin{center}
\begin{turn}{-90}
\psfrag{"twdm1keV.dat"}{Camb $ m_{FD} = 1 $ keV}
\psfrag{"2fitwdm1keV.dat"}{Fit $ m_{FD} = 1 $ keV}
\psfrag{"twdm2keV.dat"}{Camb $ m_{FD} = 2 $ keV}
\psfrag{"fitwdm2keV.dat"}{Fit $ m_{FD} = 2 $ keV}
\psfrag{"twdm4keV.dat"}{Camb $ m_{FD} = 4 $ keV}
\psfrag{"fitwdm4keV.dat"}{Fit $ m_{FD} = 4 $ keV}
\includegraphics[height=13.cm,width=8.cm]{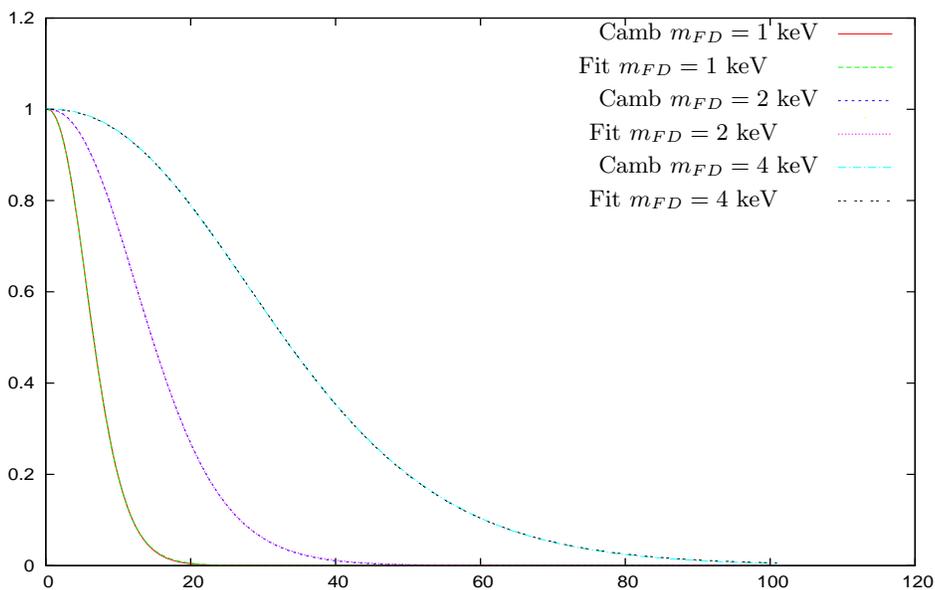}
\end{turn}
\caption{CAMB values for $ T^2(k) $ vs. $ k \; {\rm Mpc}/h $ and those from the fitting 
formula eqs.(\ref{fitt2})-(\ref{valores}) for fermionic WDM decoupling at thermal equilibrium
with masses $ m = 1 $ keV, 2 keV and 4 keV. The agreement between the fitting 
formula with the data is excellent.}
\label{fitpot}
\end{center}
\end{figure}

We display in fig. \ref{fitpot} the CAMB values for $ T^2(k) $ and the fitted
values from eqs.(\ref{fitt2})-(\ref{valores}). As one sees, the fit is excellent. 

\medskip

It is instructive to compute the wavenumber $ k_{1/2} $ where the WDM primordial power is one half
of the CDM primordial power. Namely, $ T^2(k_{1/2}) = 1/2 $. We find from eqs.(\ref{fitt2})-(\ref{valores}),
\be\label{k12}
 k_{1/2} =  \kappa \;  \left( \displaystyle 2^{\frac1{b}}-1\right)^{\frac1{a}} =
6.72 \; \left(\frac{m_{FD}}{\rm keV}\right)^{1.12} \; \frac{h}{\rm Mpc} \; .
\ee

Since $ a $ and $ b $ in eq.(\ref{fitt2}) are independent of the WDM particle mass,
we find that $ T^2(k) $ is an {\bf universal} function of $ k/k_{1/2} $ .
Therefore, we can write
\be
T^2(k) = \Phi\!\left(\frac{k}{k_{1/2}} \right) \quad , 
\ee
where
$$
\Phi(x) = \frac1{\left[1+\left(2^{1/b}-1\right) \; x^a \right]^b} \quad , \quad 2^{1/b}-1=0.167\quad ,
$$
is independent of the WDM particle mass $ m $.

\bigskip

For small and large scales the transfer function ratio $ T^2(k) $ tends to zero
and unity, respectively. These behaviours follow from eq.(\ref{fitt2}):
\bea\label{asic}
&& T^2(k) \buildrel{k \gg k_{1/2}}\over= \frac1{(2^{1/b}-1)^b} \; 
\left(\frac{k_{1/2}}{k}\right)^{a \, b} \to 0
 \quad , \cr \cr
&&T^2(k) \buildrel{k \ll k_{1/2}}\over= 1 - b \; (2^{1/b}-1) \; \left(\frac{k}{k_{1/2}} \right)^a \to 1 
\quad .
\eea
From eq.(\ref{valores}) we get the values
$$
a \; b = 10.3  \quad , \quad \frac1{(2^{1/b}-1)^b} = 2991 \quad , \quad b \; (2^{1/b}-1) = 0.748 \; .
$$

\bigskip

We see from eqs.(\ref{lfs}) and (\ref{k12}) that $ k_{1/2} < 1/l_{fs} $. 
The WDM  primordial power at $ k = 1/l_{fs} $ is strongly suppresed with respect
to the CDM primordial power. More precisely, for $ m_{FD} = 1 $ keV and $ g_d = 100 $
we find from  eqs.(\ref{lfs}) and (\ref{fitt2})-(\ref{valores}),
$$
k_{fs} = \frac1{l_{fs}} = 17.4 \; \frac1{\rm Mpc} \quad , \quad
T^2(k_{fs}) = 0.00163 \ll 1 \; .
$$
We see that $ k_{fs} $ is about four times $ k_{1/2} $ which
makes $ T^2(k = k_{fs}) $ much smaller than unity. That is, the WDM primordial power 
becomes ineffective to create structures at scales smaller than a characteristic scale below
$ l_{1/2} $ where
\be\label{lmedio}
l_{1/2} \equiv \frac1{ k_{1/2}} = 207 \; {\rm kpc} \; \left(\frac{\rm keV}{m_{FD}}\right)^{1.12} \; .
\ee
We see that $ l_{1/2} $ is practically identical to $ l_{fs} $ given by eq.(\ref{lfs}), 
(here we set $ h = 0.72 $).

\medskip

The scale $ l_{0.1}  = 1/k_{0.1} $ where $ T^2(k_{0.1}) = 0.1 $ and thus the cosmological power suppression 
is practically total for WDM, and the scale $ l_{0.9}  = 1/k_{0.9} $ where $ T^2(k_{0.9}) = 0.9 $ and
the WDM and CDM cosmological power are practically identical, are given by
\be\label{eles}
l_{0.1}  =  113 \;  {\rm kpc} \; \left(\frac{\rm keV}{m_{FD}}\right)^{1.12} \quad , \quad
l_{0.9}  = 483 \;  {\rm kpc} \; \left(\frac{\rm keV}{m_{FD}}\right)^{1.12} \quad , \quad
\frac{l_{0.9}}{l_{0.1}} = 4.27
\; .
\ee
We see from these values and $ l_{1/2} $ eq.(\ref{lmedio}),
that the WDM cosmological power decreases {\bf sharply} for decreasing scales $ l \sim l_{0.1} $.
Such sharp decrease is related to the large exponent $ a \, b \simeq 10.3 $ in the $ T^2(k) $ 
behaviour eq.(\ref{asic}).
 
\bigskip

In ref. \cite{bot} and \cite{viel05} a formula similar to eq.(\ref{fitt2})
was used to fit $ T^2(k) $ but with the constraint $ a \; b = 20 $, that is, with only two parameters
remaining free: $ \kappa $ and $ a $:
\be\label{cpma}
T^2(k) = \left[1 + (\alpha \; k)^{2 \, \nu} \right]^{-10/\nu}  \quad , \quad \nu=1.11 \; .
\ee
Fitting our CAMB results for $ T^2(k) $ with eq.(\ref{cpma})
from ref. \cite{bot} which imposes $ a \; b = 20 $, we obtain similar values for $ \alpha $ and  $ \nu $
in  eq.(\ref{cpma}) as those of ref. \cite{viel05}. This indicates that our  
CAMB results and those used in ref. \cite{viel05}
are equivalent. However, our fit to the CAMB results with our eq.(\ref{fitt2})
for $ T^2(k) $ gives a $ \chi^2 $ three times smaller than fitting the same CAMB results with 
eq.(\ref{cpma}) that imposes $ a \; b = 20 $. We conclude that our formula
eq.(\ref{fitt2}) provides a better fit than eq.(\ref{cpma}) from ref. \cite{bot}.

Notice that in our case we find from
eq.(\ref{valores})  the value $ a \; b = 10.3 $, independently of the WDM particle mass. 

\bigskip

In ref. \cite{aba1,aba2} a formula similar to eq.(\ref{fitt2}) was used to fit $ T^2(k) $
computed from CAMB for sterile neutrinos in the Dodelson-Widrow model.
The results for the coefficients of eq.(\ref{fitt2}) in ref. \cite{aba1}
are similar to ours except for their exponent $ b $ which is 35\% larger than ours in eq.(\ref{valores}).
However, our fig. \ref{fitpot} shows a better fit to the CAMB results than fig. 4 in ref. \cite{aba1}.

\subsection{WDM particles decoupling out of thermal equilibrium}\label{secnoe}

In the previous section we considered WDM particles decoupling ultrarelativistically at thermal equilibrium.

\medskip

Dark matter particles couple very weakly, weaker than weak interactions, among  themselves
and to the particles in the Standard Model of particle physics. 
This can make that they decouple at very high redshift and out of thermal equilibrium.

\medskip

Sterile neutrinos are serious WDM candidates. 
Minimal extensions of the Standard Model of particle physics 
include keV sterile neutrinos which are very weakly coupled to the standard model particles
and which are produced via the oscillation of the light (eV) active neutrinos, with their mixing
angle governing the amount of generated WDM. The mixing angle  between active and sterile neutrinos
should be in the $ \sim 10^{-4} $ scale in order to reproduce the average DM density in the
Universe.

Sterile neutrinos are usually produced out of thermal equilibrium. 
The production can be non-resonant (in the absence of lepton asymmetries) \cite{dw}
or resonantly ennhanced (if lepton asymmetries are present) \cite{sf}. In the $\nu$MSM model
sterile neutrinos are produced by the decay of a real heavier field \cite{st}.
Many particle physics models extending the Standard Model of Particle Physics
with warm dark matter particles have been proposed in the last years. These are 
models based on: Froggatt-Nielsen mechanism, flavor symmetries,
see-saw mechanisms and several variations of it, left-right symmetries
and others (see ref. \cite{merle} for a recent review).

KeV mass scale gravitinos are WDM candidates that can decouple at thermal equilibrium or
out of thermal equilibrium depending on the particle physics model \cite{fds}-\cite{gkr}.

\medskip

WDM particles decoupling ultrarelativistically in the different WDM particle models behave just as if their
masses were different \cite{fluc}. The masses of WDM particles in different models which give the same
primordial power spectrum are related according to the formula \cite{fluc}
\be\label{noneq}
m_{DW} \simeq 2.85 \; {\rm keV} \; \left(\frac{m_{FD}}{\rm keV}\right)^{4/3} \quad , \quad 
m_{SF} \simeq 2.55 \;  m_{FD}  \quad , \quad 
m_{\nu{\rm MSM}} \simeq 1.9  \; m_{FD} \; ,
\ee
where we consider the same number of internal degrees of freedom for the WDM fermions
in the different models and:

\medskip

FD stands for WDM fermions decoupling in thermal equilibrium (Fermi-Dirac).

DW  stands for WDM sterile neutrinos decoupling out of thermal equilibrium in the 
Dodelson-Widrow model \cite{dw}.

SF  stands for WDM sterile neutrinos decoupling out of thermal equilibrium in the Shi-Fuller model \cite{sf}.

$\nu$MSM  stands for WDM sterile neutrinos decoupling out of thermal equilibrium in the $\nu$MSM  model
\cite{st}.

{\vskip 0.2cm} 

The same primordial power spectrum implies identical  
differential mass function $ S(M,z) $ which we derive in sec. \ref{prsh}.

\medskip

We display in Table \ref{valK} the correspondences between the WDM particle masses in the different particle models. 

\medskip

Depending on whether the fermions be Dirac or Majorana the primordial power spectrum
is slightly  different.
Identical power spectrum follows for Dirac and Majorana fermions with masses related as \cite{fluc}
$$
m_{Maj} = 2^{1/4} \; m_{Dir} \quad {\rm in ~ the ~ FD, \; SF ~ and } \; \nu{\rm MSM \; models,}
$$
$$
m_{Maj} = 2^{1/3} \; m_{Dir} \quad {\rm in ~ the ~ DW ~ model}.
$$

\section{Bounded structures in WDM from the Press-Schechter approach. }\label{prsh}

We give in this section the mass function for WDM in the Press-Schechter approach.
For WDM studies using approaches like the halo model and excursion set to WDM see refs.\cite{mbsw}

\subsection{The expected overdensity $ \sigma(M,z) $}

As is known, from the primordial spectrum one can compute the expected 
overdensity $ \sigma(M,z) $ of mass $ M $ over a 
comoving radius $ R $ at redshift $ z $ in the linear regime. 

We consider for the DM mass $ M $:
\be\label{m}
M(R) = \frac43 \; \pi \; \Omega_{dm} \; \rho_c \;  R^3  \quad , \quad 
\sigma^2(R,z) = \int_0^{\infty} \frac{dk}{k} \; \Delta^2(k,z) \; W^2(kR) \quad , \quad h = 0.72 \quad .
\ee
$ \Delta^2(k,z) $ is the primordial power spectrum at redshift $ z $ for WDM,  $ W(x) $ is the 
 window  function:
$$
W(x)  =\frac3{x^3} \; (\sin x - x \; \cos x) \; ,
$$
and $ \rho_c $ is the critical density of the Universe.

For $ R = 8 \; {\rm Mpc}/h $ we get $ \sigma_8 = 0.797\ldots $ both for CDM and for WDM.

{\vskip 0.2cm} 

\begin{figure}[h]
\begin{center}
\begin{turn}{-90}
\psfrag{"sigmam.9.dat"}{Dodelson-Widrow}
\psfrag{"sigmam.98.dat"}{Shi-Fuller}
\psfrag{"sigmam1.3.dat"}{$\nu$MSM}
\psfrag{"sigmam2.5.dat"}{FD Thermal}
\psfrag{"sigmacdm.dat"}{CDM}
\includegraphics[height=13.cm,width=8.cm]{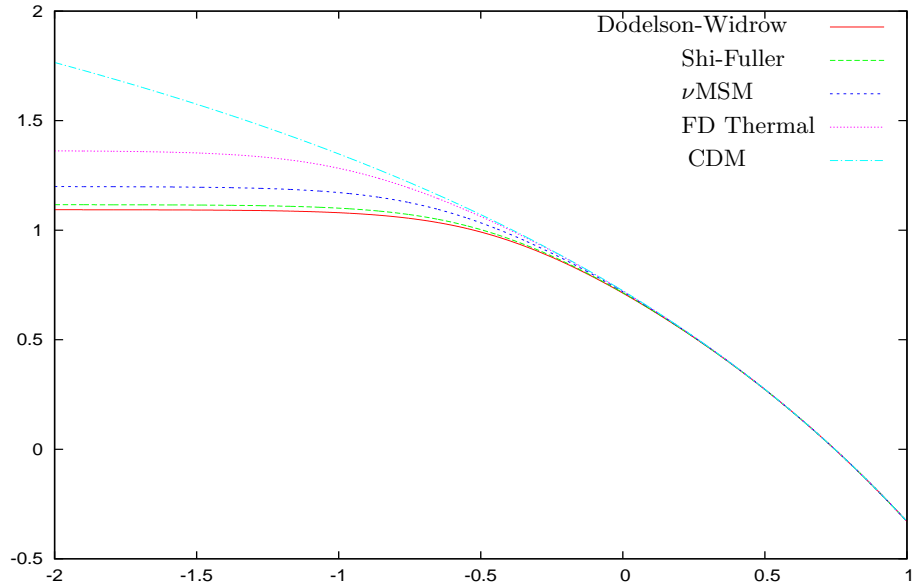}
\end{turn}
\caption{The ordinary logarithm of the expected overdensity $ \sigma^2(R,z=0) $ vs.  $ \log_{10} [R \; h / {\rm Mpc}] $ 
for a $ m =2.5 $ keV WDM particle in four different WDM models and in CDM. At small scales, $ \sigma^2(R) $ 
is almost constant for WDM. WDM flattens and reduces  $ \sigma^2(R) $ for small scales compared with CDM.}
\label{sigma}
\end{center}
\end{figure}

In fig. \ref{sigma} we plot $ \sigma(R,z=0) $ as a function of $ \log [R \; h/ {\rm Mpc}] $ for
a $ m =2.5 $ keV WDM particle in four different WDM fermion models and in CDM.

At small scales, $ \sigma^2_{WDM}(R) $ is almost constant.
WDM flattens and reduces $ \sigma^2(R) $ for small scales compared with CDM.

\medskip

It is important to characterize the linear and non-linear regimes 
in $ z $ and $ R $. The borderline between the linear 
and non-linear regimes is defined by
$$
 \sigma^2(M,z) \sim 1 
$$
Objects (galaxies) of scale $ R $ and mass $ M(R) \sim R^3 $ 
eq.(\ref{m}) start to form when the scale $ R $ becomes non-linear.

\medskip

\begin{figure}[h]
\begin{center}
\begin{turn}{-90}
\psfrag{"mnonlinm.9.dat"}{Dodelson-Widrow}
\psfrag{"mnonlinm.98.dat"}{Shi-Fuller}
\psfrag{"mnonlinm1.3.dat"}{$\nu$MSM}
\psfrag{"mnonlinm2.5.dat"}{FD Thermal}
\psfrag{"mnonlincdm.dat"}{CDM}
\includegraphics[height=13.cm,width=8.cm]{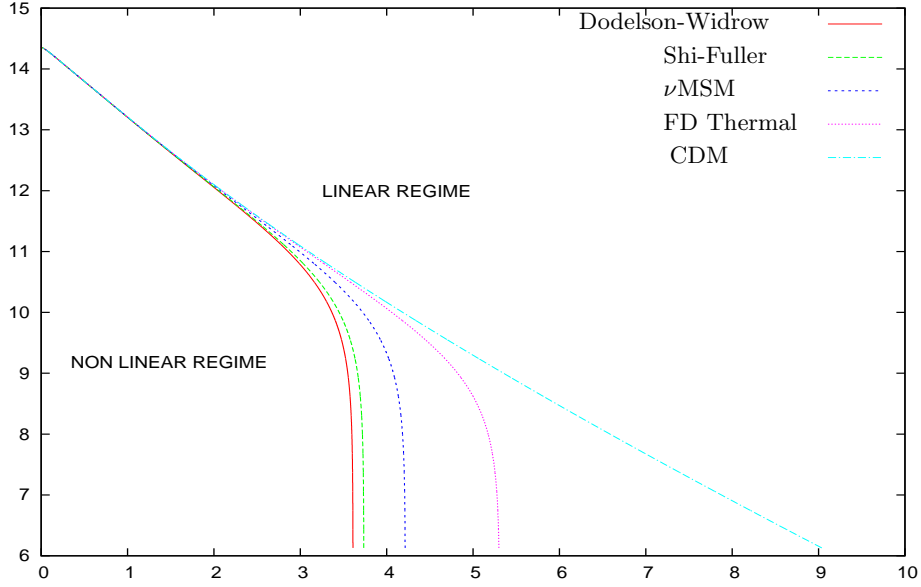}
\end{turn}
\caption{Lines where $ \sigma^2(M,z) = 1 $ in the $ z, \; \log  [ h \; M/M_\odot] $ plane
for $ m =2.5 $ keV in four different 
WDM models and in CDM. The linear regime ($ \sigma^2(M,z) < 1 $) applies above these lines 
and the nonlinear ($ \sigma^2(M,z) > 1 $) regime applies below these lines.}
\label{nolin}
\end{center}
\end{figure}

In fig. \ref{nolin} we display the lines where $ \sigma^2(M,z) = 1 $,
in the $ z, \; \log [ h \; M/{\rm Mpc}] $ plane 
for $ m =2.5 $ keV in four different WDM models and in CDM. 
Above these lines we have the linear regime where $ \sigma^2(M,z) < 1 $
and below these lines we have the nonlinear regime where $ \sigma^2(M,z) > 1 $.

We see from fig. \ref{nolin} that in WDM the linear regime lasts longer than in CDM.
That is, for a given object of mass $ M $, the nonlinear regime in WDM starts later than in CDM, 
i. e. non-linear structures start to form later in WDM than in CDM. 

In addition, in WDM, for a given particle physics model, the smaller is the particle mass,
the later the nonlinear regime starts.

We see as a general trend that decreasing the DM particle mass delays the onset of the nonlinear
regime. The earlier nonlinear onset happpens in the extreme case of heavy particles from CDM  
which masses are $ \sim $ GeV.

The dependence of the redshift where the nonlinear regime starts on the WDM particle physics model
is shown in fig. \ref{nolin} for a particle with mass $ m =2.5 $ keV.

The non-linear regime starts earlier for smaller objects than for larger ones.
Smaller objects can form earlier.

{\vskip 0.2cm} 

It is useful to compare the expected overdensities in WDM and CDM by defining the
relative overdensity $ D(R) $
\be\label{defD}
 D(R) \equiv \frac{\sigma^2_{WDM}(R,z)}{\sigma^2_{CDM}(R,z)}
\ee
Notice that the $ z $ dependence cancels out eq.(\ref{delz}) in this ratio.

\medskip

\begin{figure}[h]
\begin{center}
\begin{turn}{-90}
\psfrag{"4nsigm.9.dat"}{Dodelson-Widrow}
\psfrag{"4nsigm.98.dat"}{Shi-Fuller}
\psfrag{"4nsigm1.3.dat"}{$\nu$MSM}
\psfrag{"4nsigm2.5.dat"}{FD Thermal}
\includegraphics[height=13.cm,width=8.cm]{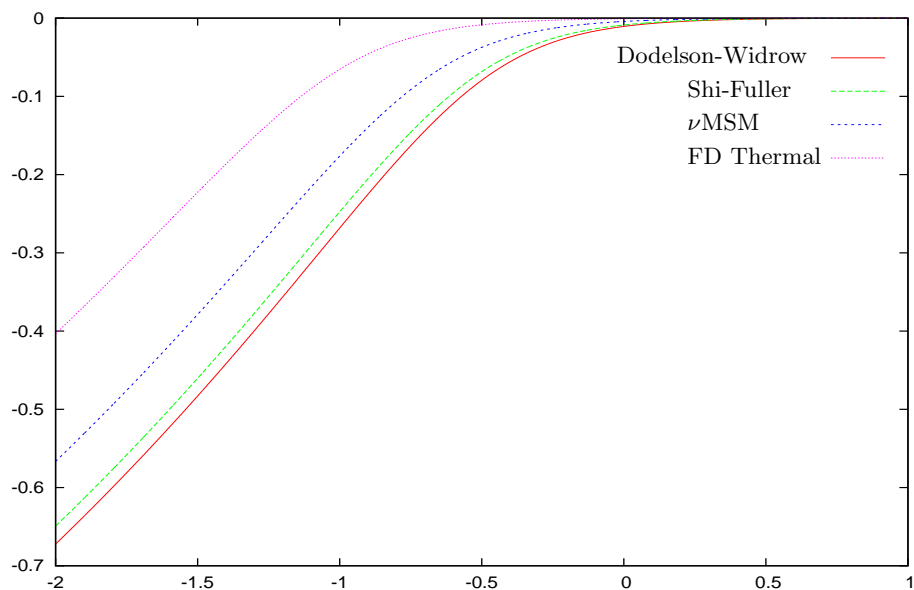}
\end{turn}
\caption{The ordinary logarithm of the relative overdensity $ D(R) = \sigma^2_{WDM}(R,z)/\sigma^2_{CDM}(R,z) $
as a function of $ \log_{10} [R \; h/{\rm Mpc}] $ for $ m =2.5 $ keV in four different WDM models.
We see in these curves the small and large scale behaviour of the relative overdensity $ D(R) $ as
described by eq.(\ref{limi}).}
\label{sigD}
\end{center}
\end{figure}

In fig. \ref{sigD} we plot $ \log_{10} D(R) $ vs. $ \log_{10} [R \; h/{\rm Mpc}] $ 
for $ m =2.5 $ keV in four different WDM models.

\medskip

It is natural to define the scale $ R_{1/2} $ where the WDM overdensity is one-half
of the CDM overdensity:
$$
 D(R_{1/2}) = \frac12 \; .
$$
For WDM particles of mass $ m_{FD} $ decoupling at thermal equilibrium we find
\be
R_{1/2} = 73.1 \; \frac{\rm kpc}{h} \; \left(\frac{\rm keV}{m_{FD}}\right)^{1.45}
\ee
Notice that $ R_{1/2} $ is about one half of $ l_{1/2} $ eq.(\ref{lmedio}).

For WDM particles decoupling out of thermal equilibrium eq.(\ref{noneq}) provides the appropriate
relationship between the masses for the different particle physics models.

{\vskip 0.2cm} 

It must be noticed that for a WDM particle mass $ m \sim 2.5 $ keV, $ R_{1/2} $ is approximately
half of $ l_{1/2} \sim  l_{fs} $.
$ l_{1/2} $ is defined by eq.(\ref{lmedio}) through the primordial power spectrum, and 
the free streaming length  $ l_{fs} $ is given by eq.(\ref{lfs}).

{\vskip 0.1cm} 

Since the overdensity is a quantity that precisely characterizes the structure formation,
one should take $ R_{1/2} $ as the typical scale below which structures are suppresed in 
WDM compared with CDM.

{\vskip 0.2cm} 

The relative overdensity $ D(R) $ vanishes in the limit $ R \to 0 $ 
reflecting the suppression of small scale structures in WDM. 
$ D(R) $ tends to unity for large scales expressing the fact that
WDM and CDM exhibit identical behaviour for large scales. 

More precisely, we find
\be\label{limi}
D(R)\buildrel{R \ll  R_{1/2}}\over= c_1 \; \left(\frac{R}{R_{1/2}}\right)^{0.37} \quad , \quad  
D(R)\buildrel{R \gg  R_{1/2}}\over= 1 - c_2 \; \left(\frac{R_{1/2}}{R}\right)^2 \; .
\ee
where $ c_1 $ and $ c_2 $ are numerical constants.

{\vskip 0.2cm} 

The relative overdensity $ D(R) $ can be reproduced by the simple formula
\be\label{fitD}
 D(R) = \frac1{\left[1 + \left(2^{1/\beta}-1\right)\left(\displaystyle
\frac{R_{1/2}}{R}\right)^\alpha\right]^\beta}
\ee
which is the configuration space analogue of eq.(\ref{fitt2}) for the power spectrum and
which reproduces the limiting behaviours in eqs.(\ref{limi}).
The values
$$
\alpha \simeq 2.2 \quad , \quad  \beta \simeq 0.17 \quad , \quad 2^{1/\beta}-1 \simeq 58 \; ,
$$
provide the best fit.

{\vskip 0.2cm} 

It is instructive to compute $ R_{0.1} $ where $ D(R_{0.1}) = 0.1 $ such that 90\% of the structures formed in
CDM are suppressed in WDM, and $ R_{0.9} $ where $ D(R_{0.9}) = 0.9 $ such that
only 10\% of the structures formed in CDM are suppressed in WDM. We find
$$
R_{0.1} = 0.980 \; \frac{\rm kpc}{h} \; \left(\frac{\rm keV}{m_{FD}}\right)^{1.45}\quad , \quad  
R_{0.9} = 495 \; \frac{\rm kpc}{h} \; \left(\frac{\rm keV}{m_{FD}}\right)^{1.45}\quad {\rm and} \quad  
\frac{R_{0.9}}{R_{0.1}} = 505 \; .
$$
We see that the scale  $ R_{0.1} $ where practically all the CDM structures are suppressed by WDM
and the scale $ R_{0.9} $ where both CDM and WDM give the same structures, are separated by 
more than two orders of magnitude: a factor 505. This follows from the smallness
of the exponent $ \simeq 0.37 $ in eq.(\ref{limi}) that governs the decrease of the 
relative overdensity $ D(R) $ for small scales.

{\vskip 0.2cm} 

Contrary to the {\bf slow} decrease of the relative overdensity $ D(R) $ for decreasing scales
$ R \sim R_{1/2} $, the transfer function $ T^2(k) $ sharply decreases in wavenumber space around $ k_{1/2} $ as
discussed in sec. \ref{sec2}. Notice that the ratio $ R_{0.9}/R_{0.1} $ is two orders of magnitude
(118 times) larger than the corresponding ratio $ l_{0.9}/l_{0.1} $  eq.(\ref{eles}).

\subsection{The differential mass function and the suppression function}

The differential mass function gives the number of isolated bounded structures 
with mass between $ M $ and $ M + dM $ per unit volume (Press-Schechter)
\be\label{ps}
\frac{dN}{dM} = - \sqrt{\frac2{\pi}} \; \frac{\delta_c}{M^2 \; \sigma^2(M,z)} \; \rho_m(z) \; 
\frac{d \sigma(M,z)}{d \ln M} \; e^{- \delta_c^2/[2 \, \sigma^2(M,z)]}  \; ,
\ee
where $ N $ is the number of structures with mass $ M $ per cubic Mpc and
the constant $ \delta_c = 1.686 \ldots $ is the linear estimate for collapse from the spherical model.

{\vskip 0.2cm} 

$ \rho_m(z) $ is the average DM density at redshift $ z $:
$$
 \rho_m(z) = \Omega_{dm} \; \rho_c \; (z+1)^3 \quad , \quad \Omega_{dm}=0.22 
\quad , \quad  \rho_c = 2.7754 \times 10^{11} \; \frac{h^2 \; M_\odot}{{\rm Mpc}^3} \quad  .
$$
{\vskip 0.2cm} 

In fig. \ref{sigma} we plot $ \sigma(M,z=0) $ vs. $ M $ for $ m =2.5 $ keV 
in the four different WDM particle models considered here and in CDM.
We see that $ \sigma(M,z) $ is almost constant in WDM for small scales and therefore
its derivative very small. The smalness of $ d\sigma(M,z)/d \ln R $ suppresses the
number of formed structures for small scales as one sees from eq.(\ref{ps}).

{\vskip 0.2cm} 

It is convenient to consider the density  $ N $  per comoving volume, namely the dimensionless
function  $ S(M,z) $
\bea\label{S}
&& \frac1{10^9 \; M_\odot \; {\rm Mpc}^3}   \; S(M,z)
   \equiv (z+1)^{-3} \; \frac{dN}{dM} \cr \cr
&& \frac1{10^9 \; M_\odot \; {\rm Mpc}^3}   \; S(M,z) = - \sqrt{\frac2{\pi}} \; 
\frac{\delta_c}{M^2 \; \sigma^2(M,z)} \; \Omega_{dm} \; \rho_c
\; \frac{d \sigma(M,z)}{d \ln M} \; e^{- \delta_c^2/[2 \, \sigma^2(M,z)]}  \; ,
\eea

{\vskip 0.1cm} 

\begin{figure}[h]
\begin{center}
\begin{turn}{-90}
\psfrag{"Rz0funmam.9.dat"}{Dodelson-Widrow}
\psfrag{"Rz0funmam.98.dat"}{Shi-Fuller}
\psfrag{"Rz0funmam1.3.dat"}{$\nu$MSM}
\psfrag{"Rz0funmam2.5.dat"}{FD Thermal}
\psfrag{"Rz0funmcdm.dat"}{CDM}
\includegraphics[height=13.cm,width=6.cm]{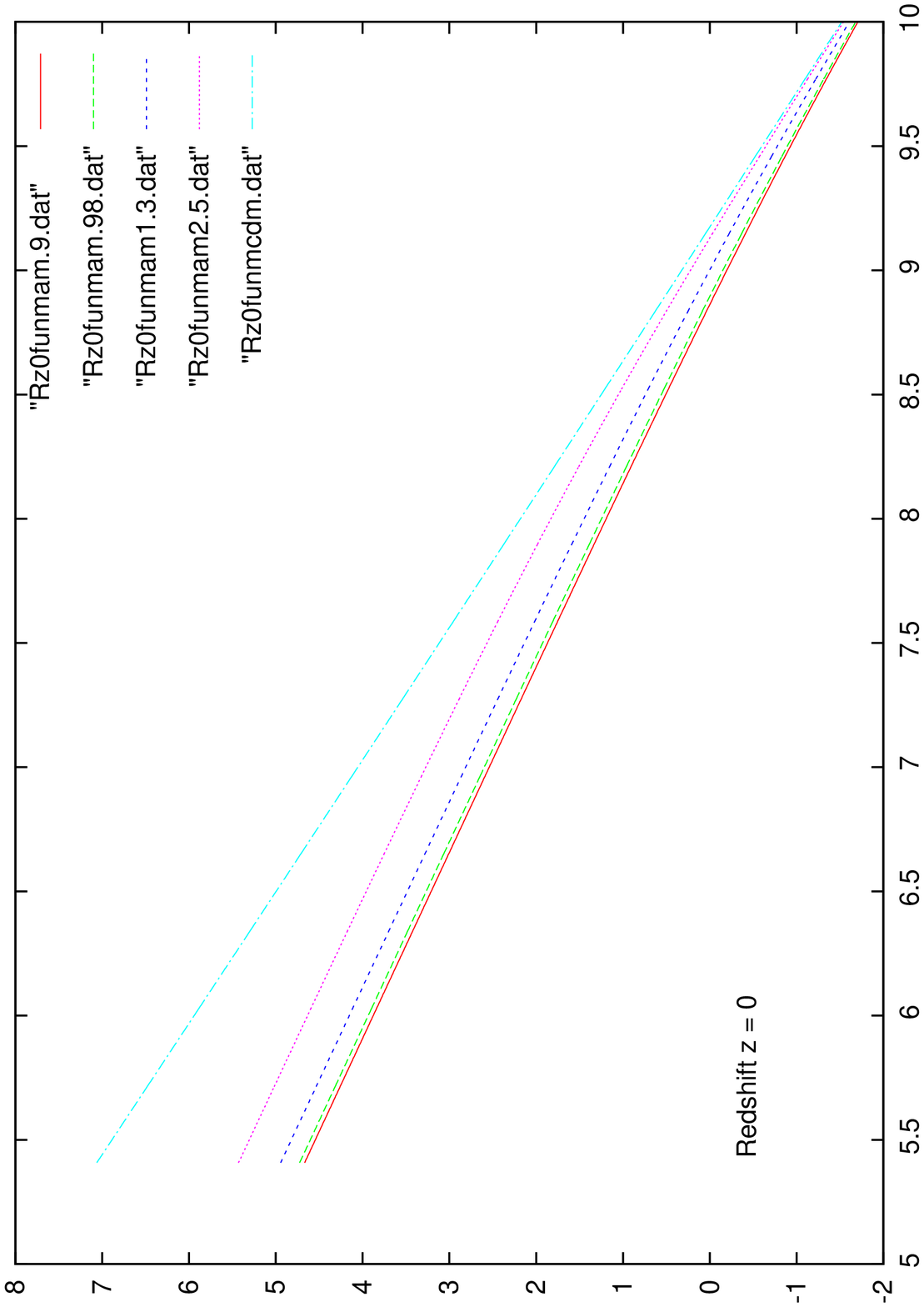}
\psfrag{"Rz2funmam.9.dat"}{Dodelson-Widrow}
\psfrag{"Rz2funmam.98.dat"}{Shi-Fuller}
\psfrag{"Rz2funmam1.3.dat"}{$\nu$MSM}
\psfrag{"Rz2funmam2.5.dat"}{FD Thermal}
\psfrag{"Rz2funmcdm.dat"}{CDM}
\includegraphics[height=13.cm,width=6.cm]{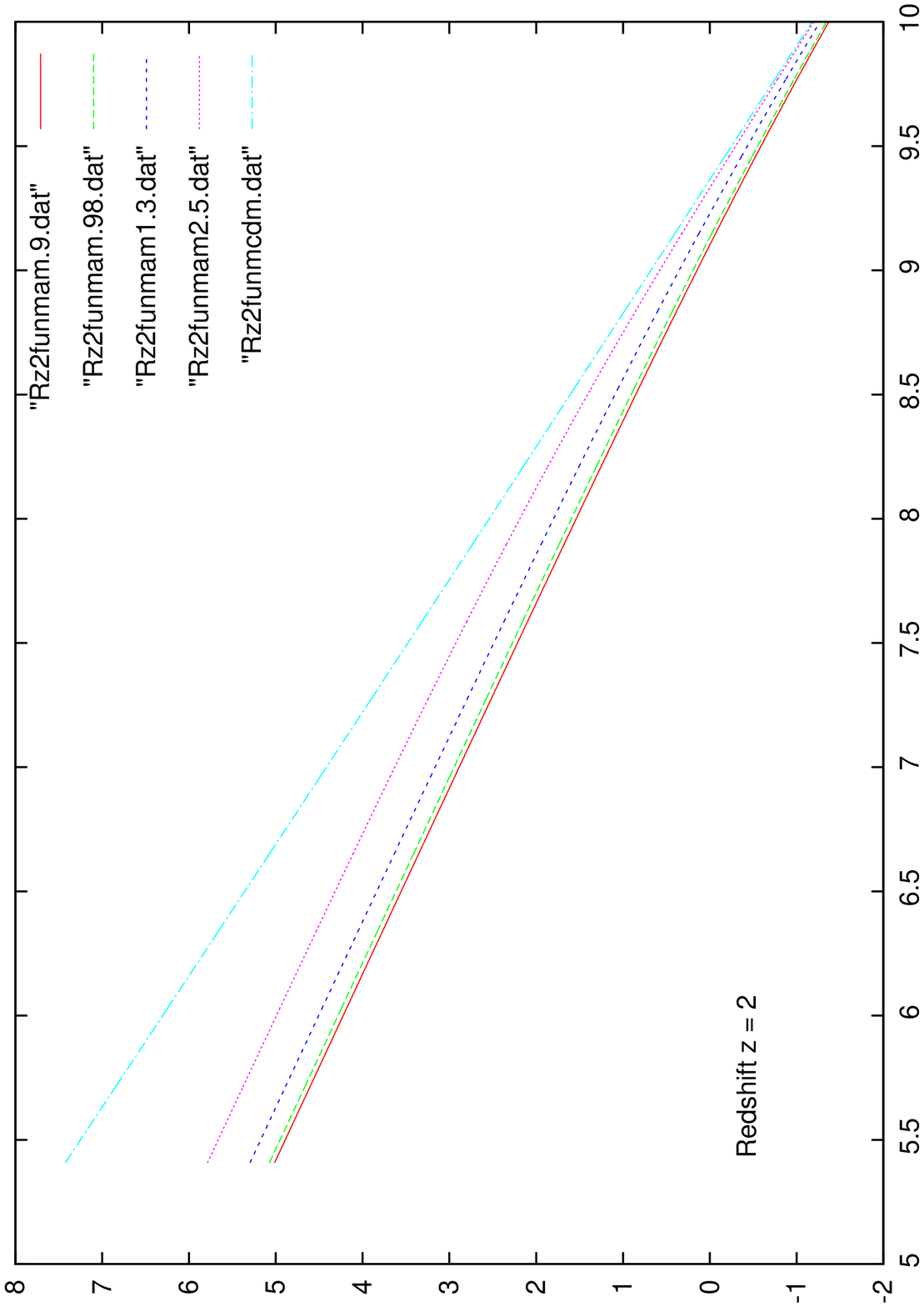}
\psfrag{"Rz5funmam.9.dat"}{Dodelson-Widrow}
\psfrag{"Rz5funmam.98.dat"}{Shi-Fuller}
\psfrag{"Rz5funmam1.3.dat"}{$\nu$MSM}
\psfrag{"Rz5funmam2.5.dat"}{FD Thermal}
\psfrag{"Rz5funmcdm.dat"}{CDM}
\psfrag{"Rcdmz5.dat"}{CDM }
\includegraphics[height=13.cm,width=6.cm]{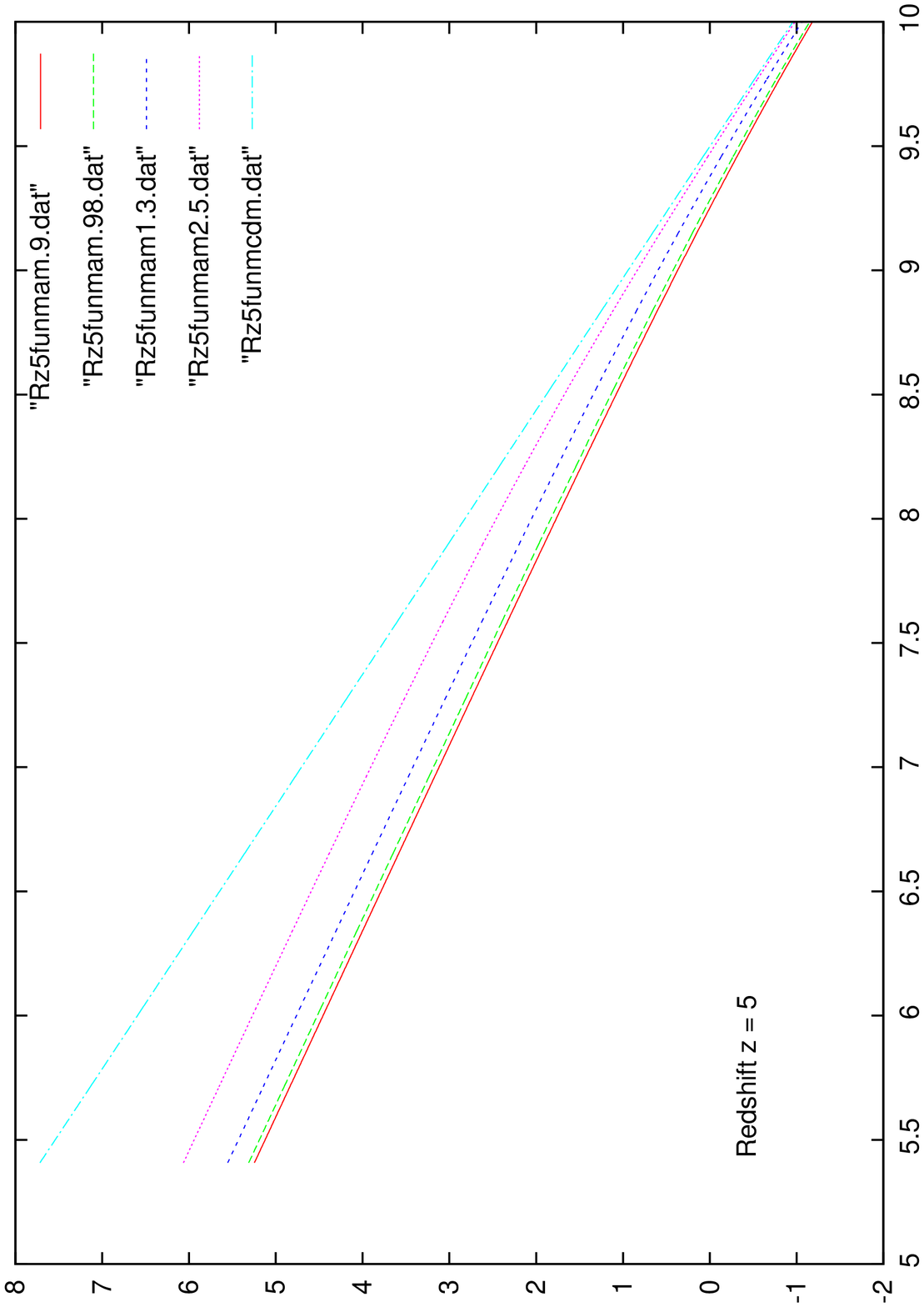}
\psfrag{"Rz10funmam.9.dat"}{Dodelson-Widrow}
\psfrag{"Rz10funmam.98.dat"}{Shi-Fuller}
\psfrag{"Rz10funmam1.3.dat"}{$\nu$MSM}
\psfrag{"Rz10funmam2.5.dat"}{FD Thermal}
\psfrag{"Rz10funmcdm.dat"}{CDM}
\includegraphics[height=13.cm,width=6.cm]{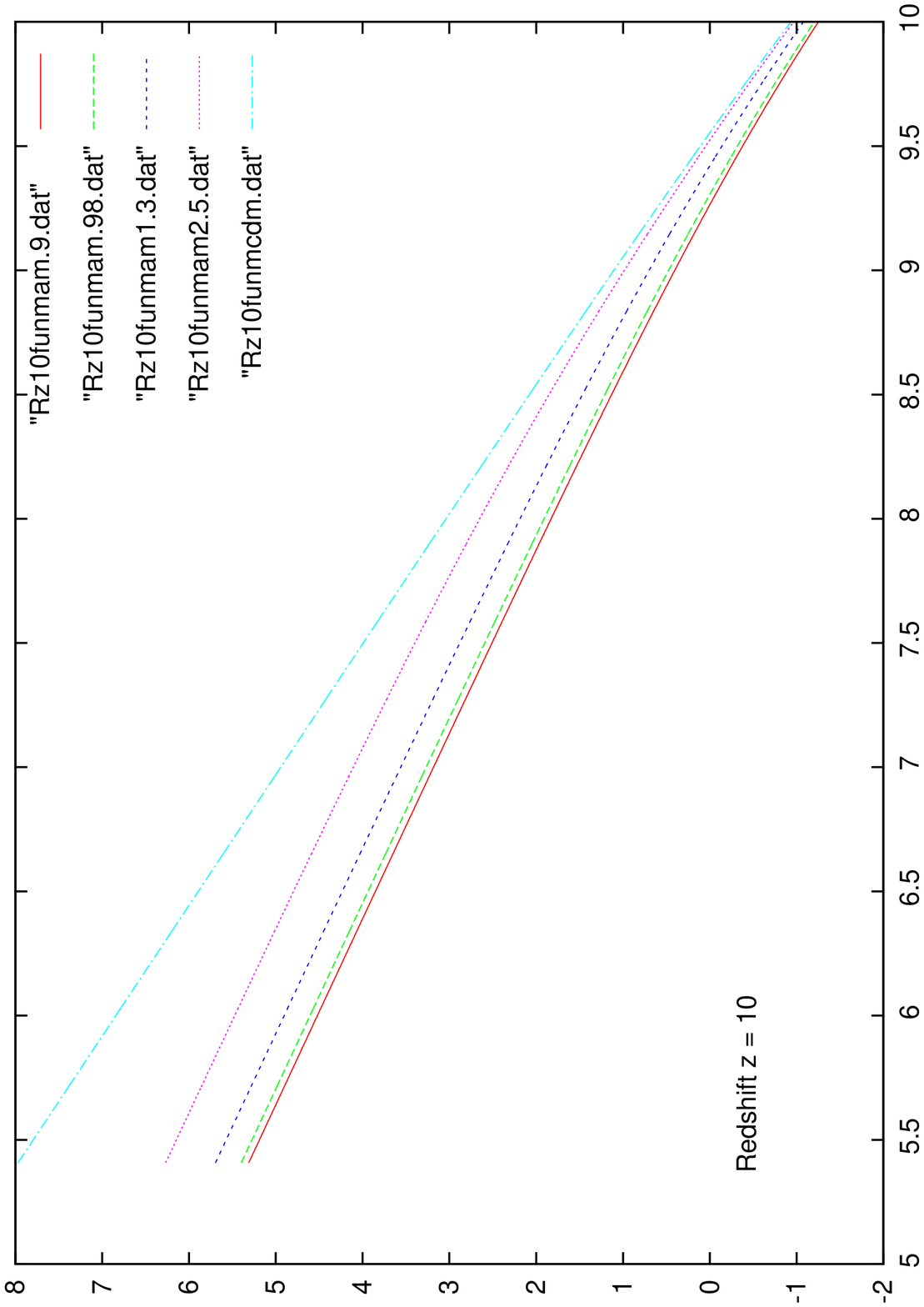}
\end{turn}
\caption{The ordinary logarithm of the dimensionless mass function  
$ \log_{10} S(M,z) $ vs. $ \log_{10} [h \; M/ M_\odot] $ at redshift 
$ z = 0, \; 2, \; 5 $ and $ 10 $ for a WDM particle of $ m = 2.5 $ keV in
four different WDM models and for CDM..}
\label{SMz}
\end{center}
\end{figure}

In figs. \ref{SMz} we plot $ \log_{10} S(M,z) $ vs. $ \log_{10} [h \; M/ M_\odot] $ 
for redshifts $ z = 0, \;  2 , \; 5 $ and $ 10 $, a WDM particle of $ m = 2.5 $ keV in
the four different WDM models and for CDM.

{\vskip 0.2cm} 

\begin{figure}[h]
\begin{center}
\begin{turn}{-90}
\psfrag{"Rz0funmam1.3.dat"}{WDM  z = 0}
\psfrag{"Rz2funmam1.3.dat"}{WDM  z = 2}
\psfrag{"Rz5funmam1.3.dat"}{WDM  z = 5}
\psfrag{"Rz10funmam1.3.dat"}{WDM  z = 10}
\psfrag{"Rz0funmcdm.dat"}{CDM  z = 0}
\psfrag{"Rz2funmcdm.dat"}{CDM  z = 2}
\psfrag{"Rz5funmcdm.dat"}{CDM  z = 5}
\psfrag{"Rz10funmcdm.dat"}{CDM  z = 10}
\includegraphics[height=13.cm,width=6.cm]{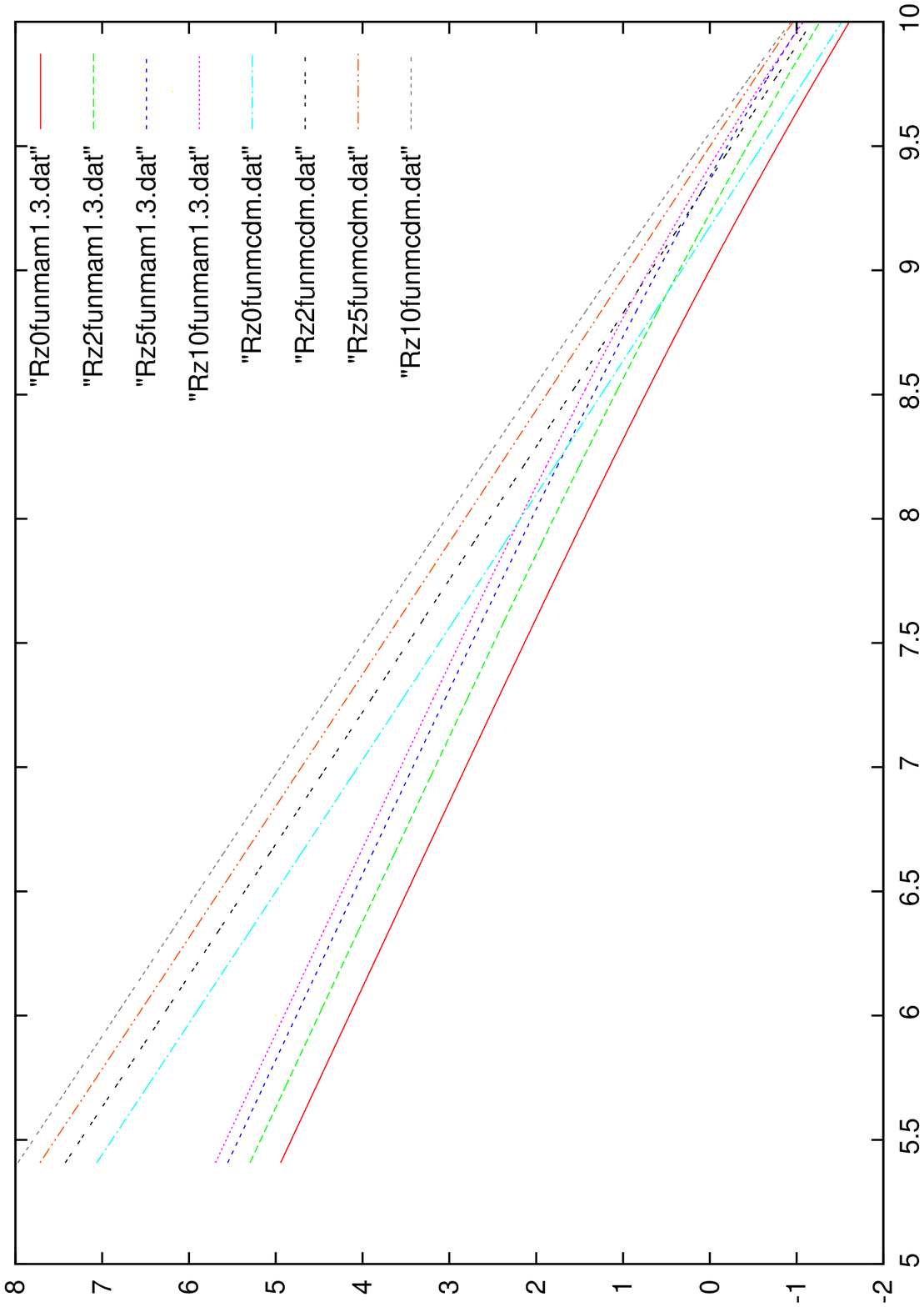}
\end{turn}
\caption{ The ordinary logarithm of the dimensionless mass function  
$ \log_{10} S(M,z) $ vs. $ \log_{10} [h \; M/ M_\odot] $ 
for a WDM particle of $ m = 2.5 $ keV in the $\nu$MSM model and in CDM
for redshifts $ z = 0, \;  2 , \; 5 $ and $ 10 $.}
\label{SMvar}
\end{center}
\end{figure}

In fig. \ref{SMvar} we plot $ \log_{10} S(M,z) $ vs. $ \log_{10} [h \; M/ M_\odot] $ 
for the a WDM particle of $ m = 2.5 $ keV in the $\nu$MSM model 
for redshifts $ z = 0, \;  2 , \; 5 $ and $ 10 $.

\begin{table}
\begin{tabular}{|c|c|c|c|} \hline  
 & &  & \\
FD (thermal)  & Dodelson-Widrow  &Shi-Fuller  & $\nu$MSM
 \\ & & & 
\\ \hline  &  &  &
   \\ 2.5 keV & 9.67 keV  & 6.38 keV & 4.75 keV \\  
 &  &  & \\ \hline  &  &  &
 \\ 0.91 keV & 2.5 keV & 2.31 keV & 1.72 keV \\  
&  &  & \\ \hline  &  &  &
\\ 0.98 keV & 2.78 keV & 2.5 keV & 1.86 keV \\ 
&  &  & \\ \hline  &  &  &
\\ 1.32 keV & 4.11 keV & 3.36 keV & 2.5 keV \\
 & & & \\ \hline  
\end{tabular}
\caption{WDM particle masses providing the same  primordial power and therefore identical  
differential mass function $ S(M,z) $ in different WDM particle models.} 
\label{valK}
\end{table}

\medskip

We see from figs. \ref{SMz} that the number of structures formed in WDM at small scales
is smaller than in CDM. As expected, the suppression of structure formation is
larger the smaller is the mass and size of the structures.

For large scales, $ M \gtrsim 10^{11} \; M_\odot $ , both CDM and WDM give the same results.

\medskip

Moreover, the suppression effect on the number of structures {\bf increases} with the redshift $ z $. 
Let us consider the suppression function $ \beta_z(M) $ introduced in eq.(\ref{defbez}) as
\be\label{betaz}
\beta_z(M) =  1 - \frac{S_{WDM}(M,z)}{S_{CDM}(M,z)} \; .
\ee
We display in Table \ref{zalto} the values of $ \beta_z(M) $ for $ z = 0 $ and
$ z = 10 $. We choose for WDM the Dodelson-Widrow model. For the other WDM particle models 
we find similar results.

{\vskip 0.1cm} 

We see in Table \ref{zalto} that for fixed redshift the suppression function $ \beta_z(M) $ decreases for
increasing $ M $ and that it increases with the redshift for fixed  $ M $. Namely, 
the suppression of structure formation in WDM compared with CDM, {\bf increases} with the redshift $ z $.

{\vskip 0.1cm} 

The suppression function $ \beta_0(M) $ at $ z = 0 $, varies from 99.1\% for masses $ 10^6 \; M_\odot/h $
to 34.8\% for masses $ 10^{10}\; M_\odot/h $.
The suppression function is {\bf higher} at $ z = 10 $ where it varies from 99.5\% to 51.8 \%
in such range of galaxy masses (Table \ref{zalto}). We see that 
for galaxy masses $ M $ about $ 10^6 \; M_\odot/h $
the suppression of structure formation in WDM does not practically change with the redshift in
the interval $ 0 < z < 10 $.
For large masses $ M = 10^{10}\; M_\odot/h $, the suppression effect at
$ z = 0 $ is 67.3\% of the suppression at $ z = 10 $ (Table \ref{zalto}).

{\vskip 0.2cm} 

Besides the variation of the suppression function with $ z $, we see in Table \ref{zalto} that in WDM,
small galaxy masses are more suppressed than large ones, as expected.

\begin{table}
\begin{tabular}{|c|c|c|c|} \hline  
&  &  &  \\
$ \frac{h \; M}{M_\odot}  $ & $ \beta_0(M) $ & $ \beta_{10}(M) $ & $  \frac{\beta_0(M)}{\beta_{10}(M)} $  
\\ & & &  \\ 
\hline  
  & & & \\ 
 $ 10^6 $ & 0.99135 & 0.99529 & 0.996 \\ 
  &  & &  \\ \hline  
  & & & \\ 
 $ 10^8 $ & 0.8973 & 0.93826 & 0.956  \\ 
  &  & &  \\ \hline  
 & & & \\ 
 $ 10^{10} $ & 0.3483 & 0.5176 & 0.673  \\ 
  &  & &  \\
\hline 
\end{tabular}
\caption{The suppression function $ \beta_z(M) $ between WDM and CDM structure formation eq.(\ref{betaz}).
The suppression of structure formation in WDM compared with CDM, {\bf turns to increase} with the redshift $ z $.
For large masses $ M = 10^{10}\; M_\odot/h $, the suppression effect at
$ z = 0 $ is 67.3\% of the suppression at $ z = 10 $.}
\label{zalto}
\end{table}

\medskip

It is therefore useful to compare these theoretical predictions with
the available observations at large $ z $.
Interestingly enough, these comparisons not only show that WDM reproduces better than CDM
the observed structures \cite{mfl} but they should point out in addition the
value of the WDM particle mass in the keV scale, and help to single out the best WDM particle physics
model.

\section{Analytic Formula for the Differential mass function}

We provide now analytic formulas for the expected overdensity
$ \sigma^2(M,z) $ that appears in the differential mass function eq.(\ref{S}).

The dependence of $ \sigma^2(M,z) $ on the redshift $ z $ factors out as 
for the primordial power spectrum eq.(\ref{delz}) \cite{wein,gor}
\be\label{Sz}
\sigma^2(M,z) = \frac1{(z+1)^2} \; \frac{g^2(z)}{g^2(0)} \; \sigma^2(M,0) \; .
\ee
We express the WDM overdensity at redshift $ z $ in terms of the CDM overdensity at redshift zero
and the relative overdensity $ D(R) $ defined by eq.(\ref{defD}),
\be\label{Sz2}
\sigma^2_{WDM}(M,z) = \frac1{(z+1)^2} \; \frac{g^2(z)}{g^2(0)} \; \sigma^2_{CDM} (M,0) \; D(R) \; .
\ee
The comoving radius $ R $ is related to the mass $ M $ by eq.(\ref{m})
\be\label{MR}
\frac{h \; R}{\rm Mpc} = 1.57540\ldots \times 10^{-4} \; \; \left(\frac{h \; M}{M_\odot}\right)^{1/3} 
\quad ,
\ee

\bigskip

At redshift $ z = 0 $, we parametrize $ \sigma^2_{CDM}(M,0) $ as
\be\label{sigR}
\sigma^2_{CDM}(M,0) = {\mathcal F}(R) = \frac{1 + x_1 \; {\check R}^2 + x_2 \; {\check R}^4}{x_3 
+ x_4 \; {\check R}^2 + x_5 \; {\check R}^4+ x_6 \; {\check R}^6}
\quad , \quad {\check R} \equiv h \; R/{\rm Mpc} \; ,
\ee
where the comoving radius $ R $ is related to the mass $ M $ by eq.(\ref{MR}) 
and $ x_i , \; i=1,...,6 $ are free parameters. 
We display in Table \ref{fit} the  parameters $ x_i $ which provide the best fit
to the numerical values for $ \sigma_{CDM}(M,0) $.

\medskip

Finally, from eqs.(\ref{Sz})-(\ref{MR}), the dimensionless differential mass function  
$ S(M,z) $ eq.(\ref{S}) can be written as 
$$
S(M,z) = 2.1698\ldots \; 10^{-4} \; \left(\frac{\rm Mpc}{h \; R}\right)^6 \; 
\; \; \frac{z+1}{g_P(z)} \; \; \displaystyle
e^{- \displaystyle \frac{0.82094\ldots \; (z+1)^2}{g_P^2(z) \; D(R) \; {\mathcal F}(R)}}
\; \; \frac{d}{d\ln R} \left[ \frac1{\sqrt{D(R) \; {\mathcal F}(R)}}\right]  
\quad , \quad h = 0.72 \; .
$$
The function $ g_P(z) $ is given by eq.(\ref{fgz}),
the function $ {\mathcal F}(R) $ by eq.(\ref{sigR}), the function $ D(R) $ by eq.(\ref{fitD})
and $ M $ and $ R $ are related by eq.(\ref{MR}). 

\begin{table}
\begin{tabular}{|c|c|c|c|c|c|} \hline  
$ x_1 $ & $ x_2 $ & $ x_3 $ & $ x_4 $ & $ x_5 $ & $ x_6 $ \\
\hline  
 148.2 & 188.7 & 0.02089 & 8.985 & 52.95 & 4.797 \\ 
\hline 
\end{tabular}
\caption{Parameters in the analytic formula eq.(\ref{sigR}) for $ \sigma^2_{CDM}(M,0) $.}
\label{fit}
\end{table}

\acknowledgments

We thank P. Salucci for useful discussions. We thank L. Danese and A. Lapi for their interest
in this work.

\appendix

\section{CDM primordial power}

We provide in this Appendix a formula that fits the primordial
power for CDM $ \Delta_{cdm}^2(k) $:
\be\label{fitcdm}
\Delta_{cdm}^2(k) = C_1 \; K^{n_s + 3} \; \left[\frac{\ln\left(1 + (q_1 \; K)^2 \right)}{(q_1 \; K)^2}\right]^2
\; \frac{1  + (q_2 \; K)^2  + (q_3 \; K)^4}{1  + (q_4 \; K)^2  + (q_5 \; K)^4} \quad , \quad 
K \equiv k \; {\rm Mpc}/h \; ,
\ee
$ C = 1.9627 \; 10^5 $ is a normalization constant, $ \; n_s = 0.96 $ stands for the spectral index.
The parameters $ q_i, \; 1 \leq i \leq 5 $
for the best fit to the CAMB data are given in Table  \ref{cdm}.

\begin{figure}[h]
\begin{center}
\begin{turn}{-90}
\psfrag{"potcdmz0.dat"}{CAMB CDM}
\psfrag{"2fitpotcdmz0.dat"}{Fit CDM}
\includegraphics[height=13.cm,width=8.cm]{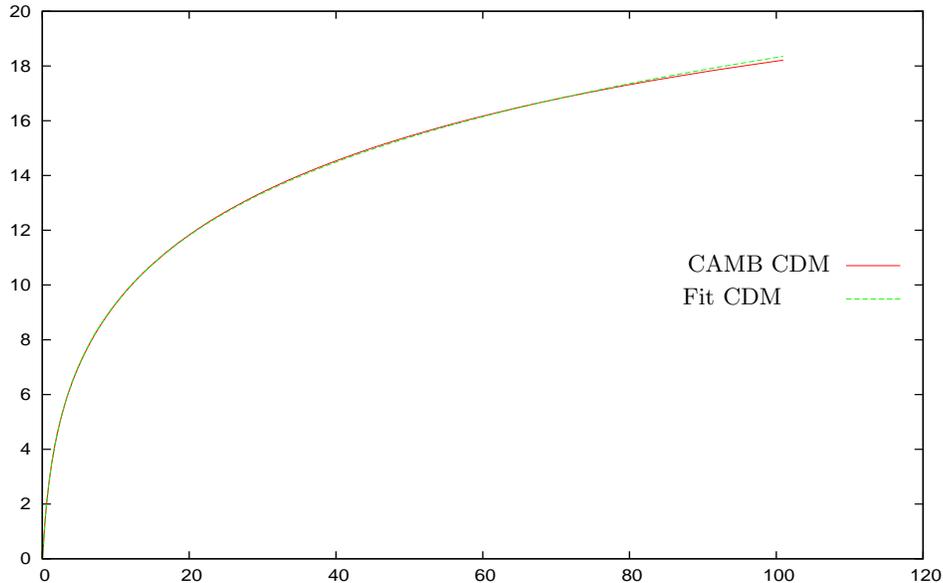}
\end{turn}
\caption{CAMB values for $ \Delta_{cdm}^2(k) $ and those from the fitting formula eq.(\ref{fitcdm})
as a function of $ k $ Mpc$/h$.}
\label{figcdm}
\end{center}
\end{figure}

We depict in fig. \ref{figcdm} the CAMB values for $ \Delta_{cdm}^2(k) $ and those from
 eq.(\ref{fitcdm}). As one can see, it is a very good fit.

A seven parameters version of the formula eq.(\ref{fitcdm}) is 
considered in ref. \cite{wein}.

{\vskip 0.2cm} 

The WDM primordial power $ \Delta_{wdm}^2(k) $ follows as the product of the CDM primordial
power $ \Delta_{cdm}^2(k) $ times the transfer function  $ T^2(k) $ according to 
eqs.(\ref{trafu}), (\ref{fitt2}) and (\ref{fitcdm})
$$
\Delta_{wdm}^2(k) =  T^2(k) \; \Delta_{cdm}^2(k) \; .
$$

\begin{table}
\begin{tabular}{|c|c|c|c|c|} \hline  
 $ q_1 $ & $ q_2 $ & $ q_3 $ & $ q_4 $ & $ q_5 $ 
\\ \hline  
13.326 & 10.525 & 5.3258 & 70.232 & 14.697  
\\ \hline  
\end{tabular}
\caption{Parameters in formula eq.(\ref{fitcdm}) for $ \Delta_{cdm}^2(k) $.}
\label{cdm}
\end{table}

\end{document}